\documentclass[12pt]{article}
\usepackage{amsmath, amsfonts, amsthm, amssymb}
\usepackage[square,sort,comma,numbers]{natbib}
\usepackage{longtable}
\usepackage[margin=1in]{geometry}
\DeclareMathOperator{\wt}{wt} 
\DeclareMathOperator{\ann}{ann} 
\DeclareMathOperator{\trace}{Tr} 

\begin{document}
\newcommand{\fpp}{\ensuremath{\mathbb{F}_{p^2}}}
\newcommand{\norm}[1]{\ensuremath{|| #1 ||}}
\newcommand{\codegenerated}[1]{\ensuremath{\langle #1 \rangle}}
\newcommand{\hip}[2]{\ensuremath{\langle #1, #2 \rangle}}
\newcommand{\conj}[1]{\ensuremath{\overline{#1}}}
\newcommand{\qecc}[4]{\ensuremath{[[#1,#2,#3]]_{#4}}}
\newcommand{\f}[1]{\ensuremath{\mathbb{F}_{#1}}}
\newcommand{\integers}{\ensuremath{\mathbb{Z}}}
\newcommand{\fzero}[1]{\ensuremath{\mathbb{F}^\star_{#1}}}
\newcommand{\hd}{\ensuremath{{\perp_h}}}
\newcommand{\todo}[1]{\textbf{TODO - } #1 }
\newcommand{\intersect}{\ensuremath{\cap}}
\newcommand{\union}{\ensuremath{\cup}}

\newcounter{thm}
\newtheorem{theorem}[thm]{Theorem}
\newtheorem*{unnum}{Theorem}
\newtheorem{observation}[thm]{Observation}
\newtheorem{lemma}[thm]{Lemma}

\title{Extending Construction X for Quantum Error-Correcting Codes}
\author{Akshay Degwekar, Kenza Guenda and T. Aaron Gulliver}
\maketitle
\begin{abstract}
In this paper we extend the work of Lisonek and Singh on
construction X for quantum error-correcting codes to finite fields
of order $p^2$ where $p$ is prime.
The results obtained are applied to the dual of Hermitian repeated root cyclic codes
to generate new quantum error-correcting codes.
\end{abstract}
\section{Introduction}
Quantum error correcting codes have been introduced as an alternative to classical
codes for use in quantum communication channels.
Since the landmark papers~\cite{shor} and~\cite{steane96}, this field of research has grown rapidly.
Classical codes have been used to construct good quantum codes~\cite{calderbank96}.
Recently, Lisonek and Singh~\cite{singh} gave a variant of Construction X that
produces binary stabilizer quantum codes from arbitrary linear codes.
In their construction, the requirement on the
duality of the linear codes was relaxed.
In this paper, we extend their work on construction X to obtain quantum error-correcting codes over finite fields of order $p^2$
where $p$ is a prime number.
We apply our results to the dual of Hermitian repeated root cyclic codes to generate new quantum
error-correcting codes.

The remainder of the paper is organized as follows.
In Section 2, we present our main result on the extension of the quantum construction X.
Section 3 characterizes the generator polynomial of the Hermitian dual of a repeated root cyclic code.
We also give the structure of cyclic codes of length $3p^s$ over
$\fpp$ as well as the structure of the dual codes.
Our interest in this class of codes comes from the importance of relaxing the
condition $(n,p)=1$, which allows us to consider codes other than the simple root codes.
\section{Extending Construction X for $\f{p}$}

Let $\f{p}$ denote the finite field with $p$ elements and
$\fzero{p} = \f{p} \backslash \{ 0\}$.
For $x \in \f{p^2}$ we denote the \emph{conjugate} of $x$ by $\conj{x} = x^p$.
Let $\hip{x}{y} = \sum^n_{i=1} x_i \conj{y_i} $ be the Hermitian inner product.
Then the \emph{norm} of $x$ is defined as $\norm{x} = \hip{x}{x} = \sum_{i=1}^n x^{p+1}$,
and the \emph{trace} of $x$ as $\trace(x) = x + \conj{x}$.
Both the trace and norm are mappings from $\fpp$ to $\f{p}$.

The following lemmas will be used later.
\begin{lemma} \label{l:existenceZ}
Let $S$ be a subspace of $\f{p^2}^n$ such that there exist $x,y$ with $\hip{x}{y} \neq 0$.
Then for all $k \in \f{p}$, there exists $z\in S$ with $\norm{z} = k$.
\end{lemma}
\begin{proof}
This is a non-constructive proof of the existence of the required element $z$.
With the assumption on $x$ and $y$, let $g(c) = \norm{cx+y} =
(cx+y)^{p+1}$ be a polynomial of degree $p+1$ in $c$.
We claim that as $c$ ranges over the elements of $\fpp$, the rhs
will range over all elements of $\f{p}$.

Assume now that there exists some $k\in \fpp$ such that $\forall c\in \fpp, g(c) \neq k$.
For each $i \in \f{p} \backslash k$, let $S_{i} = \{c \in \fpp;\, g(c)=i \}$.
Since the polynomial $g$ has degree $p+1$, $g$ can have at most $p+1$ roots in any field.
Then $|S_i| \leq p+1$, as the polynomial $g(c)-i$ can have at
most $p+1$ roots, and the $S_i$ partition the set $\fpp$.
Then $|\fpp| = p^2  \leq \sum_{i \in \f{p} \backslash {k}} |S_i| \leq (p+1)(p-1) = p^2-1$,
which is a contradiction.
Hence the result follows.
\end{proof}

\begin{lemma}\label{l:existenceB}
Let $D$ be a subspace of $\f{p^2}^n$ and assume that $M$ is a basis
for $D \intersect D^\hd$. Then there exists an orthonormal set $B$
such that $M \union B$ is a basis for $D$.
\end{lemma}
\begin{proof}
The proof given here is a generalization of the proof for the
analogous case presented in \citep[Theorem~2]{singh}.
Let $W$ be a subspace of $\f{p^2}^n$ such that
\begin{equation}
\label{e:3.1}
D = (D \intersect D^\hd ) \oplus W,
\end{equation}
and let $l = \dim(W)$.
For each $0 \leq i \leq l$, we can construct an
orthonormal set $S_i$ that is a basis for an $i$-dimensional
subspace $T_i$ of $W$ such that
\begin{equation}
\label{e:3.2Ti}
W = T_i \oplus (T_i^\hd\intersect W).
\end{equation}
The process is iterative.
Define $S_0 := \phi$ and suppose that for some $0 \leq i < l$, the set $S_i$ is an orthonormal basis for $T_i$
such that $dim(T_i) = i$ and (\ref{e:3.2Ti}) holds.
Let $x$ be a non-zero vector in $T^\hd \intersect W$.
Then there exists $y \in T^\hd\intersect W$ such that $\hip{x}{y} \neq 0$.
If no such $y$ exists, then $x\in D^\hd$, which would
contradict (\ref{e:3.1}) because the intersection of $D$ and $D^\hd$ is $\{0\}$.
Hence by Lemma \ref{l:existenceZ}, there must exist a $z \in T_i^\hd \intersect W$ such that $\norm{z} = 1$.
Set $S_{i+1}=S_i\union \{z\}$.
Clearly all the elements in $S_{i+1}$ are orthogonal to each other.
In addition, $\norm{s} = 1$ for all $s\in S_{i+1}$.

Let $T_{i+1}$ be the subspace spanned by $S_{i+1}$.
As $z \not \in T_{i}$ we have that $\dim(T_{i+1}) = i+1$.
To show that
\begin{equation}
\label{3.1Ti+1} W = T_{i+1} \oplus (T_{i+1}^\hd \intersect W),
\end{equation}
we must first show that $T_{i+1}\intersect T_{i+1}^\hd \intersect W = {0}$.
Let $v \in T_{i+1} \intersect T_{i+1}^\hd \intersect W$.
As $v \in T_{i+1}$, we have $v = u + cz$ where $u \in T_i$ and $c \in \f{p^2}$.
Since $v \in T_{i+1}^\hd$, we have for each $w \in T_i$
and each $d \in \f{p^2}$ that
\begin{equation*}
0 = \hip{u + cz}{w + dz} = \hip{u}{w} + \conj{d}\hip{u}{z} +
c\hip{z}{w} + c\conj{d}\norm{z} = \hip{u}{w} + c\conj{d}.
\end{equation*}
We must have $c = 0$ or else $\hip{u}{w} + cd$ would not remain
constant as $d$ runs over the elements of $\fpp$.
Thus $\hip{u}{w}=0$ for all $w \in T_{i}$, and hence $u \in T_{i}^\hd$. As $u \in T_i$ and $T_i
\intersect T_{i}^\hd = {0}$, we obtain that $u=0$.
Hence $v$ is also $0$ and $T_{i+1}\intersect T_{i+1}^\hd \intersect W = {0}$.

Next we show that $W = T_{i+1}  + (T_{i+1} \intersect W)$.
Let $w \in W$.
By assumption $W = T_i + (T_i^\hd \intersect W)$, so
there exist vectors $x \in T_i$ and $y \in T_i^\hd \intersect W$
such that $w = x + y$.
Now it is shown that $W=T_{i+1}+(T_{i+1}^{\hd}\intersect W)$.
By assumption $W=T_{i}+(T_{i}^{\hd}\bigcap W)$, so there
exist vectors $x\in T_{i}$ and $y\in T_{i}\intersect W$.
Clearly $x\in T_{i+1}$ and for any $u+dz\in T_{i+1}$ (where $u\in
T_{i}$ and $d\in\fpp$), we have
\begin{eqnarray}
\hip{y-\hip{y}{z}z}{u+dz} &=& \hip{y}{u}+\conj{d}\hip{y}{z}-\hip{y}{z}\hip{z}{u}-\conj{d}\hip{y}{z}\norm{z} \nonumber \\
&=&\conj{d}\hip{y}{z}-\conj{d}\hip{y}{z} \nonumber \\
&=&0.
\end{eqnarray}
Thus $y\in T_{i+1}\intersect W$, and hence $W=T_{i+1}+(T_{i+1}\intersect W)$.
This completes the proof that (\ref{e:3.2Ti}) implies
(\ref{3.1Ti+1}) assuming that the vector $z$ is chosen as described above.
\end{proof}

\begin{theorem}
\label{th:main}
For an $[n,k]_{p^2}$ linear code $C$, let $e = n-k-\dim(C \cap C^\hd)$.
Then there exists a quantum code with parameters
$\qecc{n+e}{2k-n}{d}{p}$ with $d \geq \min(\wt(C), \wt(C + C^\hd)+1)$.
\end{theorem}
\begin{proof}
We start with the observation that the equation $x^2+1=0$ always has
a solution in $\fpp$. This can be proven using the fact that
$\fpp^\star$ is a cyclic group. Let $\beta $ be a generator of $\fpp^*$.
Then $\beta^k = -1$ for some $k$, and it is also known that $-1^2 = 1$.
Hence $\beta^{2k}=1$ and $p^2-1 | 2k$, so that $k$ is even.
Thus, $\beta^\frac{k}{2}$ is the required solution.

As defined previously
\[
e=\dim(C^{\hd})-\dim(C\text{\ensuremath{\cap}}C^{\hd})=\dim(C+C^{\hd})-\dim(C).
\]
Let $s=\dim(C\cap C^{\hd})$, and $G$ be the matrix
\begin{equation}
\label{m:generatorG}
G=\begin{pmatrix}M_{s\times n} & 0_{s\times e}\\
A_{(n-e-2s)\times n} & 0_{(n-e-2s)\times e}\\
B_{e\times n} & \beta^{k/2}I_{e\times e}
\end{pmatrix},
\end{equation}
where the size of the matrix is indicated by the subscripts, and
$0$ and $I$ denote the zero matrix and identity matrix, respectively.

For a matrix $P$, let $r(P)$ denote the set of rows of $P$.
The matrix $G$ is constructed such that $r(M)$ is a basis for $C\cap
C^{\hd}$, $r(M)\cup r(A)$ is a basis for $C$, $r(M)\cup r(B)$ is a
basis for $C$, and $r(B)$ is an orthonormal set.
The existence of such a matrix $B$ follows from Lemma \ref{l:existenceB}.
Note that $r(M)\cup r(A)\cup r(B)$ is a basis for $C+C^{\hd}$ .

Let $E$ be the linear code for which $G$ is a generator matrix.
Further, let $S$ denote the union of the first $s$ rows of $G$ and the last $e$
rows of $G$, i.e., $S$ is the set of rows of the matrix
\begin{equation}
\label{m:generatorS}
S=\begin{pmatrix}M_{s\times n} & 0_{s\times e}\\
B_{e\times n} & \beta^{k/2}I_{e\times e}
\end{pmatrix}.
\end{equation}
We observe that each row of $S$ is orthogonal to each row of $G$
because any row from the first $s$ rows of $S$ represents a vector in
$C\cap C^{\hd}$, and hence is orthogonal with all codewords in $C+C^{\hd}$,
the code represented by $G$.

Consider a row from the last $e$ rows in $S$. This row is orthogonal
to the first $n-e-s$ rows of $G$ because they represent the code $C$
while the matrix $B$ represents codewords from $C^{\hd}$. The rows
of the matrix are orthogonal. Because in the case they are different
rows in the matrix, then they are orthogonal and the $\beta^{k/2}I$
matrix part will contribute a $0$. Any row $z$ is self-orthogonal
since from the construction $\norm{z}=1$ and the identity matrix
will contribute $-1$, giving an inner product of $0$. This completes
the proof of the observation. Thus, each vector from $S$ belongs to
$E^{\hd}$, and the vectors in $S$ are linearly independent because
\[
\dim(E^{\hd})=n+e-(n-s)=e+s=|S|.
\]
Hence $S$ is a basis for $E^{\hd}$.
Since $S$ is a subset of $G$ by construction, it follows that $E^{\hd}\subseteq E$.

Let $x$ be a non-zero vector in $E$ and due to the
vertical block structure of $G$, we can write $x=(x^{1}|x^{2})$
where $x^{1}\in\fpp^{n}$ and $x^{2}\in\fpp^{e}$.
Thus $x$ is a linear combination of rows of $G$.
If none of the last $e$ rows of $G$ are contained in this linear combination with a non-zero coefficient,
then $x^{1}\in C\backslash{0}$, and so $\wt(x)=\wt(x^{1})\ge\wt(C)$.
If some of the last $e$ rows of $G$ are in this linear combination with
a non-zero coefficient, then $x^{1}\in C+C^{\hd}$ and
$\wt(x)=\wt(x^{1})+\wt(x^{2})\ge\wt(C+C^\hd)+1$.
Thus $E$ is an $[n+e,k+e,d]_{p^{2}}$ code with
$d\ge\min(\wt(C),\wt(C+C^{\hd})+1)$ and $E^{\hd}\subseteq E$.
The code $E$ satisfies the required conditions, and thus the proof is complete.
\end{proof}

Many constructions of quantum codes use self-orthogonal
codes~\cite{G2010,G-G2013}, which corresponds to the case when $e=0$.
The results of the next section are required to construct the quantum codes in subsequent sections.
Note that many of the results in the next section can easily be generalized to constacyclic codes.

\section{The Hermitian Dual of Repeated Roots Cyclic Codes}
Let $p$ is a prime number and $C$ be a cyclic code of length $n$ over the finite field $\fpp$.
Then $C$ is given by the principal ideal $g(x)$ in $\dfrac{\fpp[x]}{\langle x^{n}-1 \rangle}$,
and so $g(x)$ is called the generator polynomial for $C$.
When the length $n$ divides $p$, $C$ is called a repeated root cyclic code.

In this section, we obtain the generator polynomial of the Hermitian
dual of a repeated root cyclic code.
We also give the structure of the cyclic codes of length $3p^s$ over $\fpp$ as well as the
structure of the dual code.
Our interest in this class of codes comes from the importance of relaxing the condition $(n,p)=1$,
which allows us to consider codes other than simple root codes.

Let $f(x)=a_0+a_1 x+\ldots + a_rx^r$ be a polynomial in
$\mathbb{F}_{q^2}[x]$, and $\conj{f(x)} = \conj{a_0} + \conj{a_1}x + \ldots + \conj{a_r}x^r$.
The polynomial inverse of $f$ is denoted by $f^\star(x) = x^rf(x^{-1}) = a_r+a_{r-1} x+\ldots + a_0x^r$,
so then $f^{\bot}(x) = \overline{a_r} + \overline{a_{r-1}}x + \ldots + \overline{a_0}x^r$.

The following properties can easily be verified.

\begin{lemma}
\label{l:propConjInv}
Let $f(x)$ and $g(x)$ be polynomials over $\f{p^m}$.
Then
\begin{enumerate}
\item conjugation is additive: $\conj{f(x) + g(x)} =\conj{f(x)} + \conj{g(x)}$;
\item conjugation is multiplicative: $\conj{f(x)g(x)} =\conj{f(x)}\, \conj{g(x)}$;
\item polynomial inversion is additive if the polynomials have the same degree:\\
${(f(x) + g(x))}^\star ={f(x)}^\star + {g(x)}^\star $;
\item polynomial inversion is multiplicative: ${(f(x)g(x))}^\star ={f(x)}^\star \,{g(x)}^\star$;
\item inversion and conjugation commute with each other: $\conj{(f(x)^\star)} = (\conj{f(x)})^\star$; and
\item both operations are self-inverses: $(f(x)^\star)^\star = f(x)$ and $\conj{\conj{f(x)}} = f(x)$.
\end{enumerate}
\end{lemma}
\begin{lemma}
\label{l:herDualCondition1} Let $a(x) = a_0 + a_1x + \ldots + a_{n-1}x^{n-1}$
and
$b(x) = b_0 + b_1x + \ldots + b_{n-1}x^{n-1}$ be polynomials in $\dfrac{F_{p^2}[x]}{x^n-1}$.
Then $a(x)\conj{b(x)} = 0$ in $\dfrac{F_{p^2}[x]}{x^n-1}$
if and only if $(a_0, a_1, \ldots, a_{n-1})$ is orthogonal to
$(\conj{b_{n-1}}, \conj{b_{n-2}}, \ldots, \conj{b_0})$ and all its cyclic shifts.
That is $\hip{a}{\conj{b^\star}}=0  \iff a(x)b(x)^\bot =0$.
\end{lemma}
\begin{proof}
It well known (see for example~\cite{huffman03}), that if $a(x) = a_0 + a_1x + \ldots + a_{n-1}x^{n-1}$
and $b(x) = b_0 + b_1x + \ldots + b_{n-1}x^{n-1}$ are
polynomials in $\dfrac{F_{p^2}[x]}{x^n-1}$, then $a(x)b(x) = 0$ in
$\dfrac{F[x]}{x^n-1}$ if and only if $(a_0, a_1, \ldots, a_{n-1})$
is orthogonal to $(b_{n-1}, b_{n-2},\ldots, b_0)$ and all its cyclic shifts.
Hence by applying this fact to $a(x)$ and
$\conj{b(x)}$ and noting that $\conj{\conj{b(x)}} = b(x)$, the result follows.
\end{proof}

We now use Lemma~\ref{l:herDualCondition1} to derive an expression for the Hermitian dual
of a cyclic code.
Let $S\subseteq R$ and let the annihilator be $\ann(S) = \{ g\in R | fg=0, \; \forall f\in S\}$.
Then $\ann(S)$ is also an ideal of the ring and hence is generated by a polynomial.
\begin{lemma}
\label{l:hdToAnnihilator}
If $g(x)$ generates the code $C$, then $C^\hd = \ann(\conj{g(x)}^\star)$.
\end{lemma}
\begin{proof}
Assume that $g(x)$ generates the code $C$.
Then each codeword in $C$ has the form  $a(x) = g(x)c(x)$.
Let a codeword $b(x)$ lie in the Hermitian dual $C^\hd$.
Then by Lemma \ref{l:herDualCondition1} we have that
\[
     a(x)b^\bot(x)=0,
\]
and by Lemma \ref{l:propConjInv}, this is equivalent to
\begin{equation}
\label{l:annihilatorDerivation}
 b(x) (\conj{g(x)}^\star) =0.
\end{equation}
Then by (\ref{l:annihilatorDerivation}), we have that for a
codeword $b(x)$, $b(x)\in C^\hd \iff b(x)\in \ann(\conj{g(x)}^\star)$,
which completes the proof.
\end{proof}

\begin{lemma}
\label{l:annihilatorToGeneratorPoly}
Assume that $C=\codegenerated{g(x)} $ is a cyclic code of length $n$ over
$\mathbb{F}_{p^2}$ with generator polynomial $g(x)$.
Define $h(x)=\frac{x^n-1}{g(x)}$.
Then we have that $C^{\hd}= \codegenerated{h^{\bot}(x)}$.
\end{lemma}
\begin{proof}
From Lemma \ref{l:hdToAnnihilator} it is known that $C^\hd = \ann(g(x)^\bot)$.
Thus, we must show that $\ann(g^\bot(x)) = \codegenerated{h^\bot(x)}$.
One way containment is easy since $\codegenerated{h^\bot(x)} \subseteq \ann(g^\bot(x))$,
which is true because $h^\bot(x)g^\bot(x) = (h(x)g(x))^\bot = (x^n-1)^\bot = 0$ by Lemma \ref{l:propConjInv}.
For containment the other way, we observe that since
$\ann(g^\bot(x))$ is an ideal of the polynomial ring
$\dfrac{\fpp[x]}{x^n-1}$, it is generated by a polynomial, say
$b^\bot(x)$. Then $b^\bot(x)g^\bot(x) = x^n -1 =
\lambda(x^n-1)^\bot$ (because $b(x)$ is the smallest polynomial, so it is an equality).
Hence $b(x)g(x) = x^n -1$, so it must be that
$b(x)= h(x)$ since both are unitary polynomials.
This completes the proof.
\end{proof}

\begin{theorem}
Let $p > 3$ be a prime.
Then
\begin{enumerate}
\item there exists $\omega \in \fpp$ such that $\omega^3=1$ and the factorization of $x^{3p^s}-1$ into irreducible factors over $\fpp[x]$ is
\[
x^{3p^s}-1 = (x-1)^{p^s}(x-\omega)^{p^s}(x-\omega^2)^{p^s};
\]
\item the cyclic codes of length $3p^s$ are always of the form
\[ \codegenerated{(x-1)^{i}(x-\omega)^{j}(x-\omega^2)^{k}},\] where $0
\leq i,j,k \leq p^s$, and the code has $p^{2(3p^s-i-j-k)}$ codewords; and
\item the Hermitian dual of the codes have the form
\begin{equation} \label{e:hdGeneratorForm}
C^\hd =
    \begin{cases}
        \codegenerated{(x-1)^{p^s-i}(x-\omega)^{p^s-j}(x-\omega^2)^{p^s-k}} & \text{ if } p\equiv 1\mod 3, \\
        \codegenerated{(x-1)^{p^s-i}(x-\omega^2)^{p^s-j}(x-\omega)^{p^s-k}} & \text{ if } p\equiv 2 \mod 3.
    \end{cases}
\end{equation}
\end{enumerate}
\end{theorem}
\begin{proof}
    \begin{enumerate}
    \item Since $p$ is a prime number, $p \neq 0 \mod 3$, and $p^2-1 = (p+1)(p-1)$, so either
     $p+1 = 0 \mod 3$ or  $p-1 = 0 \mod 3$. Therefore an element of order 3 exists in $\fpp$.
     Let this element be $\omega$, so then $(x-1)(x-\omega)(x-\omega^2) = x^3-1$.
    In a field of characteristic $p$, it is known that $x^n-1 = (x^m-1)^p$ if $n=mp$.
    Therefore we have that $x^{3p^s}-1 = (x^3-1)^{p^s} = ((x-1)(x-\omega)(x-\omega^2))^{p^s}.$
    \item
        From the previous part we know that the irreducible factors are $(x-1)$, $(x-\omega)$ and $(x-\omega^2)$,
        each of multiplicity $p^s$.
        As the generator polynomial divides $x^{3p^s}-1$, the statement follows.
    \item
        We know from Lemma \ref{l:annihilatorToGeneratorPoly} that
        \[C^\hd = \codegenerated{h^\bot(x)}, \]
hence
\begin{align}
C^\hd &= \conj{ \codegenerated{\dfrac{(x-1)^{p^s}(x-\omega)^{p^s}(x-\omega^2)^{p^s}}{(x-1)^{i}(x-\omega)^{j}(x-\omega^2)^{k}}}^\star } \nonumber \\
    &= \conj{ \codegenerated{(x-1)^{p^s-i}(x-\omega)^{p^s-j}(x-\omega^2)^{p^s-k}}^\star } \nonumber \\
    &= \conj{\codegenerated{[(x-1)^{p^s-i}]^\star[(x-\omega)^{p^s-j}]^\star [(x-\omega^2)^{p^s-k}]^\star} }  \nonumber \\
    &= \conj{\codegenerated{[-(x-1)^{p^s-i}][- \omega(x-\omega^{-1})^{p^s-j}]^\star [- \omega^2 (x-\omega^{-2})^{p^s-k}]^\star} } \nonumber \\
    & \text{Since, } (x-1)^\star = -x + 1 = -(x-1) , (x-\omega)^\star = -\omega x+1 = -\omega (x-\omega^2) \nonumber \\
    &= \conj{\codegenerated{[(x-1)^{p^s-i}][(x-\omega^2)^{p^s-j}][(x-\omega)^{p^s-k}]} } \nonumber \\
    &= \codegenerated{[(\conj{x-1})^{p^s-i}][(\conj{x-\omega^2})^{p^s-j}][(\conj{x-\omega})^{p^s-k}]}  \nonumber \\
    &= \codegenerated{[({x-1})^{p^s-i}][({x-\omega^{2p}})^{p^s-j}][({x-\omega^p})^{p^s-k}]}  \nonumber \\
    &= \begin{cases}
        \codegenerated{(x-1)^{p^s-i}(x-\omega^2)^{p^s-j}(x-\omega)^{p^s-k}} & \text{ if } p\equiv 1\mod 3, \\
        \codegenerated{(x-1)^{p^s-i}(x-\omega)^{p^s-j}(x-\omega^2)^{p^s-k}} & \text{ if } p\equiv 2 \mod 3.
    \end{cases}          \\
    & \text{Since } \omega^{p} = \omega \text{ if } p \equiv 1 \mod 3 \text{, and } \omega^{p} = \omega^2 \text{ if } p \equiv 2 \mod 3.
\end{align}
    \end{enumerate}
This completes the proof.
\end{proof}

\section{Extension to Simple Root Cyclic Codes}
This section considers cyclic codes of length $n$ over $\fpp$ such that $(p,n)=1$.
In this case, a cyclic code can be represented by its defining set $Z$.
If $m$ has order $p^{2}$ modulo $n$, then $\f{p^{2m}}$ is the splitting field of
$x^n-1$ containing a primitive $n$th root of unity.
Consider a primitive root $\beta$.
Then $\{ k|g(\beta^{k})=0,\; 0\leq k<n \}$ is a defining set of $C$.
Note that this set depends on the choice of $\beta$.
We can make a canonical choice for $\beta$ by fixing a primitive element $\alpha$ of $\f{q^m}$ and
letting $\beta=\alpha^{\frac{q^{m}-1}{n}}$.
Let this elements be $\alpha$ as defined by the \textsf{PrimitiveElement} function in Magma.

For $n$ and $m$ as above and $a\in\{0, \ldots, n-1\}$, the set $\{aq^j
\mod n | 0 \leq j < m \}$ is called a \emph{cyclotomic coset modulo} $n$.
It is well known that a defining set of a cyclic code of length $n$
is the union of \emph{cyclotomic cosets modulo} $n$.
Let $\integers_{n}$ denote the set of integers modulo $n$.
Clearly defining sets can be considered as subsets of $\integers_{n}$.
For $S \subset \integers_n$, denote $\conj{S}=\integers_{n}
\backslash S$ and $-qS = \{-qs \mod n | s \in S \}$.

We now prove the following lemma.
\begin{lemma}
If $C$ is a linear cyclic code with defining set $Z$, then
$\dim(C^\hd) - \dim (C \intersect C^\hd)  = |Z \cap -pZ|$.
\end{lemma}
\begin{proof}
Let $C$ be a linear cyclic code of length $n$, and $\prod_{k\in
Z}(x-\beta^{k})$ be the generator polynomial for $C$.
Then from
Lemma~\ref{l:annihilatorToGeneratorPoly} the generator polynomial
for $C^{\hd}$ is $\prod_{k\in-p\conj{Z}}(x-\beta^{k})$, and the
generator polynomial for $C\cap C^{\hd}$ is $\prod_{k\in
Z\cap-p\conj{Z}}(x-\beta^{k})$, which gives that
\[
\dim(C^{\hd})-dim(C\cap C^{\hd})=n-|-p\conj{Z}|-(n-|Z\cup-p\conj{Z}|)=|Z\cup-p\conj{Z}|-|-p\conj{Z}|=|Z\cap-pZ|.
\]
\end{proof}

\begin{theorem}
Assume $n$ be divisible by $p^{2}-1$ and let $C$ be an $[n,k]_{4}$ cyclic
code with defining set $Z$ such that
$(Z\cap-pZ) \subseteq T = \{\frac{nk}{p^{2}-1}|k\in\{1,\ldots, p^{2}-1\}\}$.
If $e=|Z\cap-pZ|$, then there exists an $[[n+e,2k-n+e,d]]_{p}$ quantum
code with $d\geq\min\{\wt(C),\wt(C_{u})+1,\wt(C+C^{\hd})+2\}$ where
the minimum is taken over the cyclic codes $C_{u}$ with defining set
$Z\backslash \{u \}$ for each $u\in Z\cap-pZ$.
\end{theorem}
\begin{proof}
The proof requires a modification to the proof of Theorem \ref{th:main}, in particular
the set of orthonormal vectors used is changed.
First we observe that each of the elements in $T$ is a cyclotomic
coset and contains only one element.
Let $q = p^2-1$, $n = (p^2-1)l = ql$, and $\omega$ be a $p^2-1$-th root of unity.
Consider the polynomials
\[
b_t(x) = \dfrac{x^n-1}{x-\omega^t} = \sum_{i=0}^{l-1} (x^{qi+q-1} + \omega^t x^{qi+q-2} + \ldots + \omega^{(q-1)t}x^{qi}).\]
For convenience, we let $\{b_i | i\in {0,1,\dots, l} \}$ also denote the corresponding codewords.
This is an orthonormal set because
\[
\hip{b_u}{b_v} = q\sum_{i=0}^{l-1} (\omega^{i (u + vp) }) =q\sum{i=0}^{l-1} (\omega^{i (u - v) }) = \begin{cases} 0 & u \neq v \\
                                    ql & u = v \end{cases}.
\]
To mitigate the $ql$ factor, we can multiply each element by a constant.
Thus, to add the rows for $B$ to the matrix, we add $U=\{b_t | \frac{tn}{q} \in Z \cap -pZ \}$.

To prove the claim about the distance, we have 3 cases: no row from
B is a linear combination, exactly one row from $U$ is a linear combination
with a non-zero coefficient, and at least two rows are a combination.
The proof of the first and the last cases is the same as in the proof of Theorem \ref{th:main}.
For the second case, let $b_t$ be the row with non-zero coefficient.
Then the code generated would be $span(C, b_t)$, which is precisely the cyclic code with defining set
$Z \backslash \{\frac{tn}{3}\}$.
This completes the proof.
\end{proof}

\section{Examples of Codes Generated}
In this section, a comprehensive table of codes generated is
presented. Many of these codes have parameters better than the best
known quantum codes. 
\begin{longtable}{| p{.25\textwidth} | p{.50\textwidth} |  p{.25\textwidth}|}
\hline
New Codes & Generator Polynomial& Best Known Binary QECC \\
\hline $[[33, 31, 2]]_{3}$ & $x^{13} + \alpha^5*x^{12} +
\alpha^7*x^{11} + \alpha^2*x^{10} + 2*x^9 + 2*x^8 + \alpha^3*x^7 +
\alpha^6*x^6 + 2*x^5 + \alpha^3*x^4 + \alpha^3*x^3 + \alpha^6*x^2 +
\alpha^2$
 &$[[33, 31, 1]]_2$ \\
$[[35, 33, 2]]_{3}$ & $x + 1$ &$[[35, 33, 1]]_2$ \\
$[[39, 37, 2]]_{3}$ & $x+2$ &$[[39, 37, 1]]_2$ \\
$[[40, 26, 5]]_{3}$ & $x^7 + \alpha*x^6 + \alpha*x^5 + \alpha^6*x^4 + x^3 + \alpha^7*x^2 + \alpha^5*x + \alpha$ &$[[40, 26, 4]]_2$ \\
$[[40, 24, 6]]_{3}$ & $x^8 + \alpha^3*x^7 + \alpha*x^6 + \alpha^7*x^5 + 2*x^4 + x^3 + \alpha^2*x^2 + \alpha*x + \alpha^2$ &$[[40, 24, 5]]_2$ \\
$[[41, 39, 2]]_{3}$ & $x + 1$ &$[[41, 39, 1]]_2$ \\
$[[41, 9, 11]]_{3}$ & $x^{16} + \alpha^5*x^{15} + \alpha^5*x^{14} + \alpha*x^{13} + \alpha^6*x^{12} + \alpha^2*x^{11} + 2*x^9 + \alpha^7*x^7 + 2*x^6 + \alpha^3*x^5 + 2*x^4 + \alpha^6*x^3 + \alpha^7*x^2 + \alpha^7$ &$[[41, 9, 8]]_2$ \\
$[[41, 19, 8]]_{3}$ & $x^{11} + \alpha*x^{10} + \alpha*x^9 + \alpha^5*x^8 + 2*x^7 + \alpha^3*x^6 + \alpha^2*x^5 + \alpha^3*x^4 + x^2 + \alpha^2*x + 2$ &$[[41, 19, 6]]_2$ \\
$[[40, 20, 7]]_{3}$ & $x^{10} + \alpha^{6}*x^{9} + \alpha^{7}*x^{6} + \alpha^{3}*x^{5} + {2}*x^{4} + \alpha^{3}*x^{3} + \alpha*x + \alpha^{2}$ &$[[40, 20, 6]]_2$ \\
$[[41, 25, 6]]_{3}$ & $x^{8} + \alpha^{6}*x^{7} + \alpha^{7}*x^{6} + x^{4} + \alpha^{5}*x^{3} + \alpha^{3}*x^{2} + {2}*x + \alpha^{7}$ &$[[41, 25, 4]]_2$ \\
$[[40, 10, 10]]_{3}$ & $x^{15} + \alpha^{6}*x^{14} + x^{13} +
\alpha^{5}*x^{{12}} + \alpha^{2}*x^{11} + x^{10} + \alpha^{3}*x^{9}
+ \alpha^{5}*x^{8} + x^{7} + {2}*x^{6} + \alpha^{7}*x^{5} +
\alpha^{7}*x^{4} + x^{3} + \alpha*x^{2} + \alpha^{5}*x + \alpha^{5}
$ &$[[40, 10, 8]]_2$ \\
$[[40, 16, 8]]_{3}$ & $x^{12} + \alpha^{3}*x^{11} +
\alpha^{3}*x^{10} + x^{9} + \alpha^{5}*x^{8} + \alpha^{5}*x^{7} +
\alpha^{5}*x^{6} + \alpha^{5}*x^{5} + {2}*x^{4} + x^{3} +
\alpha^{6}*x + \alpha^{2}
$ &$[[40, 16, 6]]_2$ \\
$[[41, 13, 9]]_{3}$ & $x^{14} + {2}*x^{13} + \alpha*x^{12} +
{2}*x^{10} + \alpha^{2}*x^{9} + x^{8} + \alpha^{5}*x^{7} +
\alpha^{5}*x^{6} + x^{5} + \alpha^{3}*x^{4} + \alpha^{6}*x^{3} +
{2}*x^{2} + \alpha^{3}*x + \alpha^{5}
$ &$[[41, 13, 7]]_2$ \\
$[[41, 21, 7]]_{3}$ & $x^{10} + \alpha^{7}*x^{9} + \alpha^{6}*x^{7} + \alpha^{6}*x^{6} + {2}*x^{5} + \alpha^{7}*x^{4} + x^{3} + \alpha^{3}*x + \alpha^{5}$ &$[[41, 21, 6]]_2$ \\
$[[41, 27, 5]]_{3}$ & $x^{7} + {2}*x^{6} + \alpha^{3}*x^{5} + \alpha^{7}*x^{4} + \alpha^{7}*x^{3} + {2}*x^{2} + \alpha^{3}*x + \alpha^{2}$ &$[[41, 27, 4]]_2$ \\
$[[33, 31, 2]]_{3}$ & $ x + {1}$ &$[[33, 31, 1]]_2$ \\
$[[40, 26, 5]]_{3}$ & $x^{7} + \alpha*x^{6} + \alpha*x^{5} + \alpha^{6}*x^{4} + x^{3} + \alpha^{7}*x^{2} + \alpha^{5}*x + \alpha$ &$[[40, 26, 4]]_2$ \\
$[[40, 24, 6]]_{3}$ & $x^{8} + \alpha^{3}*x^{7} + \alpha*x^{6} + \alpha^{7}*x^{5} + {2}*x^{4} + x^{3} + \alpha^{2}*x^{2} + \alpha*x + \alpha^{2}$ &$[[40, 24, 5]]_2$ \\
$[[41, 39, 2]]_{3}$ & $ x + {1}$ &$[[41, 39, 1]]_2$ \\
$[[41, 9, 11]]_{3}$ & $x^{16} + \alpha^{5}*x^{15} + \alpha^{5}*x^{14} + \alpha*x^{13} + \alpha^{6}*x^{12} + \alpha^{2}*x^{11} + {2}*x^{9} + \alpha^{7}*x^{7} + {2}*x^{6} + \alpha^{3}*x^{5} + {2}*x^{4} + \alpha^{6}*x^{3} + \alpha^{7}*x^{2} + \alpha^{7}$ &$[[41, 9, 8]]_2$ \\
$[[41, 19, 8]]_{3}$ & $x^{11} + \alpha*x^{10} + \alpha*x^{9} +
\alpha^{5}*x^{8} + {2}*x^{7} + \alpha^{3}*x^{6} + \alpha^{2}*x^{5} +
\alpha^{3}*x^{4} + x^{2} + \alpha^{2}*x + {2}
$ &$[[41, 19, 6]]_2$ \\
$[[40, 20, 7]]_{3}$ & $x^{10} + \alpha^{6}*x^{9} + \alpha^{7}*x^{6} + \alpha^{3}*x^{5} + {2}*x^{4} + \alpha^{3}*x^{3} + \alpha*x + \alpha^{2}$ &$[[40, 20, 6]]_2$ \\
$[[41, 25, 6]]_{3}$ & $x^{8} + \alpha^{6}*x^{7} + \alpha^{7}*x^{6} +
x^{4} + \alpha^{5}*x^{3} + \alpha^{3}*x^{2} + {2}*x + \alpha^{7}
$ &$[[41, 25, 4]]_2$ \\
$[[40, 10, 10]]_{3}$ & $x^{15} + \alpha^{6}*x^{14} + x^{13} +
\alpha^{5}*x^{12} + \alpha^{2}*x^{11} + x^{10} + \alpha^{3}*x^{9} +
\alpha^{5}*x^{8} + x^{7} + {2}*x^{6} + \alpha^{7}*x^{5} +
\alpha^{7}*x^{4} + x^{3} + \alpha*x^{2} + \alpha^{5}*x + \alpha^{5}
$ &$[[40, 10, 8]]_2$ \\
$[[40, 16, 8]]_{3}$ & $x^{12} + \alpha^{3}*x^{11} + \alpha^{3}*x^{10} + x^{9} + \alpha^{5}*x^{8} + \alpha^{5}*x^{7} + \alpha^{5}*x^{6} + \alpha^{5}*x^{5} + {2}*x^{4} + x^{3} + \alpha^{6}*x + \alpha^{2}$ &$[[40, 16, 6]]_2$ \\
$[[41, 13, 9]]_{3}$ & $x^{14} + {2}*x^{13} + \alpha*x^{12} +
{2}*x^{10} + \alpha^{2}*x^{9} + x^{8} + \alpha^{5}*x^{7} +
\alpha^{5}*x^{6} + x^{5} + \alpha^{3}*x^{4} + \alpha^{6}*x^{3} +
{2}*x^{2} + \alpha^{3}*x + \alpha^{5}
$ &$[[41, 13, 7]]_2$ \\
$[[41, 21, 7]]_{3}$ & $x^{10} + \alpha^{7}*x^{9} + \alpha^{6}*x^{7} + \alpha^{6}*x^{6} + {2}*x^{5} + \alpha^{7}*x^{4} + x^{3} + \alpha^{3}*x + \alpha^{5}$ &$[[41, 21, 6]]_2$ \\
$[[41, 27, 5]]_{3}$ & $x^{7} + {2}*x^{6} + \alpha^{3}*x^{5} + \alpha^{7}*x^{4} + \alpha^{7}*x^{3} + {2}*x^{2} + \alpha^{3}*x + \alpha^{2}$ &$[[41, 27, 4]]_2$ \\
$[[41, 9, 11]]_{5}$ & $x^{16} + \alpha*x^{15} + \alpha^{23}*x^{14} + \alpha^{3}*x^{13} + {4}*x^{12} + \alpha^{15}*x^{11} + {3}*x^{10} + \alpha^{10}*x^{8} + \alpha^{3}*x^{7} + \alpha^{14}*x^{5} + \alpha^{8}*x^{4} + \alpha^{19}*x^{3} + {4}*x^{2} + \alpha^{17}*x + \alpha^{8}$ &$[[41, 9, 8]]_2$ \\
$[[40, 2, 12]]_{5}$ & $x^{19} + \alpha^{20}*x^{18} + \alpha^{22}*x^{17} + \alpha^{10}*x^{16} + \alpha^{3}*x^{15} + \alpha^{20}*x^{14} + \alpha^{21}*x^{13} + \alpha^{22}*x^{12} + \alpha^{20}*x^{11} + \alpha^{8}*x^{10} + {3}*x^{9} + \alpha^{22}*x^{8} + {4}*x^{7} + \alpha^{11}*x^{6} + \alpha^{23}*x^{5} + \alpha^{22}*x^{4} + \alpha^{8}*x^{3} + \alpha^{5}*x^{2} + \alpha^{9}*x + \alpha^{4}$ &$[[40, 2, 10]]_2$ \\
$[[41, 5, 12]]_{5}$ & $x^{18} + \alpha^{15}*x^{17} +
\alpha^{19}*x^{16} + \alpha^{23}*x^{15} + \alpha^{13}*x^{14} +
\alpha^{5}*x^{13} + \alpha^{7}*x^{12} + x^{11} + \alpha^{17}*x^{10}
+ \alpha^{3}*x^{9} + \alpha^{19}*x^{8} + \alpha^{19}*x^{7} +
\alpha^{5}*x^{6} + \alpha^{11}*x^{5} + \alpha^{8}*x^{4} +
\alpha*x^{3} + \alpha^{10}*x^{2} + \alpha^{5}*x + {1}
$ &$[[41, 5, 9]]_2$ \\
$[[40, 6, 11]]_{5}$ & $x^{17} + \alpha^{10}*x^{16} +
\alpha^{4}*x^{15} + \alpha^{22}*x^{14} + \alpha^{9}*x^{13} +
\alpha*x^{12} + {3}*x^{11} + {4}*x^{10} + \alpha^{5}*x^{9} +
\alpha^{16}*x^{8} + \alpha^{19}*x^{7} + \alpha^{22}*x^{6} +
\alpha^{9}*x^{5} + \alpha^{4}*x^{4} + {4}*x^{3} + \alpha^{17}*x^{2}
+ \alpha^{16}*x + {4}
$ &$[[40, 6, 8]]_2$ \\
$[[39, 15, 9]]_{5}$ & $x^{12} + {2}*x^{11} + \alpha^{5}*x^{10} +
\alpha^{16}*x^{9} + \alpha^{3}*x^{8} + \alpha^{3}*x^{7} +
\alpha^{13}*x^{6} + \alpha^{15}*x^{5} + \alpha^{22}*x^{4} + x^{3} +
\alpha^{9}*x^{2} + {2}*x + \alpha^{16}
$ &$[[39, 15, 7]]_2$ \\
$[[39, 23, 5]]_{5}$ & $x^{8} + \alpha^{21}*x^{7} + {3}*x^{6} + \alpha*x^{5} + \alpha^{16}*x^{4} + \alpha^{17}*x^{3} + \alpha^{2}*x^{2} + \alpha^{21}*x + \alpha^{16}$ &$[[39, 23, 4]]_2$ \\
$[[40, 22, 6]]_{5}$ & $x^{9} + \alpha^{7}*x^{8} + \alpha^{8}*x^{7} + \alpha^{2}*x^{6} + \alpha^{21}*x^{5} + \alpha^{9}*x^{4} + \alpha^{14}*x^{3} + \alpha^{20}*x^{2} + \alpha^{19}*x + {4}$ &$[[40, 22, 5]]_2$ \\
$[[41, 21, 7]]_{5}$ & $x^{10} + \alpha^{3}*x^{9} + x^{8} + \alpha^{10}*x^{7} + \alpha^{2}*x^{6} + \alpha^{22}*x^{5} + \alpha^{23}*x^{4} + \alpha*x^{3} + \alpha^{22}*x^{2} + \alpha^{15}*x + {1}$ &$[[41, 21, 6]]_2$ \\
$[[39, 11, 10]]_{5}$ & $x^{14} + \alpha^{15}*x^{12} + \alpha^{21}*x^{11} + \alpha^{16}*x^{10} + \alpha^{16}*x^{9} + {4}*x^{8} + \alpha^{3}*x^{7} + {4}*x^{5} + {4}*x^{4} + \alpha^{22}*x^{3} + \alpha^{19}*x^{2} + \alpha^{9}*x + \alpha^{8}$ &$[[39, 11, 8]]_2$ \\
$[[39, 19, 7]]_{5}$ & $x^{10} + \alpha^{14}*x^{8} + \alpha^{14}*x^{7} + \alpha^{4}*x^{6} + \alpha*x^{5} + {4}*x^{4} + \alpha^{8}*x^{3} + \alpha^{3}*x^{2} + \alpha^{14}*x + \alpha^{8}$ &$[[39, 19, 5]]_2$ \\
$[[40, 18, 8]]_{5}$ & $x^{11} + x^{10} + \alpha^{13}*x^{9} + \alpha^{17}*x^{8} + {2}*x^{7} + \alpha^{14}*x^{6} + \alpha^{17}*x^{5} + {3}*x^{4} + \alpha^{15}*x^{3} + \alpha^{21}*x^{2} + \alpha^{23}*x + {4}$ &$[[40, 18, 6]]_2$ \\
$[[31, 13, 6]]_{5}$ & $x^{9} + {3}*x^{8} + x^{6} + x^{5} + {4}*x^{4} + x^{3} + {3}*x^{2} + x + {4}$ &$[[31, 13, 5]]_2$ \\
$[[32, 0, 11]]_{5}$ & $x^{16} + {3}*x^{15} + {2}*x^{14} + x^{13} + x^{11} + {2}*x^{10} + x^{9} + x^{8} + {4}*x^{7} + x^{6} + x^{5} + {3}*x^{4} + {2}*x^{3} + x + {1}$ &$[[32, 0, 10]]_2$ \\
$[[31, 7, 8]]_{5}$ & $x^{12} + {4}*x^{11} + {4}*x^{10} + {2}*x^{9} + {4}*x^{8} + {2}*x^{7} + x^{6} + {3}*x^{5} + x^{4} + x^{3} + {2}*x + {1}$ &$[[31, 7, 7]]_2$ \\
$[[32, 12, 7]]_{5}$ & $x^{10} + {3}*x^{7} + x^{6} + x^{5} + x^{4} + {3}*x^{2} + {4}*x + {1}$ &$[[32, 12, 6]]_2$ \\
$[[31, 25, 3]]_{5}$ & $ x^{3} + x^{2} + {3}*x + {4}$ &$[[31, 25, 2]]_2$ \\
$[[32, 18, 5]]_{5}$ & $x^{7} + {3}*x^{5} + {3}*x^{3} + {4}*x^{2} + {4}$ &$[[32, 18, 4]]_2$ \\
$[[32, 6, 9]]_{5}$ & $x^{13} + {2}*x^{11} + x^{10} + x^{9} +
{4}*x^{8} + {3}*x^{6} + {2}*x^{5} + {4}*x^{3} + {4}*x^{2} + {4}*x +
{4}
$ &$[[32, 6, 8]]_2$ \\
$[[33, 31, 2]]_{5}$ & $x + {4}$ &$[[33, 31, 1]]_2$ \\
$[[35, 33, 2]]_{5}$ & $x + {4}$ &$[[35, 33, 1]]_2$ \\
$[[37, 35, 2]]_{5}$ & $x + \alpha^{16}$ &$[[37, 35, 1]]_2$ \\
$[[37, 35, 2]]_{5}$ & $x + \alpha^{16}$ &$[[37, 35, 1]]_2$ \\
$[[25, 23, 2]]_{5}$ & $x + \alpha^{16}$ &$[[25, 23, 1]]_2$ \\
$[[24, 20, 3]]_{5}$ & $x^{2} + \alpha^{8}*x + \alpha^{17}$ &$[[24, 20, 2]]_2$ \\
$[[25, 21, 3]]_{5}$ & $x^{2} + \alpha^{13}*x + \alpha^{17}
$ &$[[25, 21, 2]]_2$ \\
$[[24, 18, 4]]_{5}$ & $x^{3} + \alpha^{10}*x^{2} + \alpha^{16}*x +
{3}
$ &$[[24, 18, 2]]_2$ \\
$[[25, 19, 4]]_{5}$ & $x^{3} + \alpha^{19}*x^{2} + \alpha^{10}*x + \alpha^{21}$ &$[[25, 19, 2]]_2$ \\
$[[25, 17, 5]]_{5}$ & $x^{4} + \alpha^{7}*x^{3} + {4}*x^{2} + \alpha^{16}*x + {3}$ &$[[25, 17, 3]]_2$ \\
$[[33, 31, 2]]_{5}$ & $x + {4}$ &$[[33, 31, 1]]_2$ \\
$[[32, 0, 13]]_{7}$ & $x^{16} + {2}*x^{15} + {3}*x^{14} + {4}*x^{12} + x^{11} + {4}*x^{10} + x^{9} + {5}*x^{8} + {4}*x^{7} + {6}*x^{6} + {5}*x^{5} + {2}*x^{4} + {3}*x^{3} + {3}*x^{2} + {4}*x + {1}$ &$[[32, 0, 10]]_2$ \\
$[[31, 1, 12]]_{7}$ & $x^{15} + {3}*x^{14} + {6}*x^{13} + {6}*x^{12} + {3}*x^{11} + {4}*x^{10} + x^{9} + {2}*x^{8} + {4}*x^{6} + {3}*x^{5} + x^{4} + {3}*x^{3} + {6}*x^{2} + {2}*x + {6}$ &$[[31, 1, 11]]_2$ \\
$[[33, 21, 5]]_{7}$ & $x^{6} + \alpha^{42}*x^{5} + \alpha^{33}*x^{4} + \alpha^{20}*x^{3} + \alpha^{30}*x^{2} + \alpha^{6}*x + \alpha^{15}$ &$[[33, 21, 4]]_2$ \\
$[[33, 31, 2]]_{7}$ & $x + {6}$ &$[[33, 31, 1]]_2$ \\
\hline
\caption{Comparison of the codes obtained using Theorem 3 presented here and the best known binary QECC} 
\label{tab:AgainstBinaryQECC}
\end{longtable}

\section{Codes Generated from Repeated Root Cylic Codes}
%
 Codes of length $3p^s$ on field of size $p^2$.
\begin{longtable}{| p{.33\textwidth} | p{.33\textwidth} |  p{.34\textwidth}|}
\hline
 Codes & Codes & Codes \\ \hline
$[[15,9,2]]_{25}$ & $[[15,7,3]]_{25}$ & $[[16,6,4]]_{25}$ \\
$[[75,69,2]]_{25}$ &$[[75,59,3]]_{25}$ &$[[75,49,4]]_{25}$ \\
$[[82,26,5]]_{25}$ & & \\
$[[375,369,2]]_{25}$ & $[[375,319,3]]_{25}$ &$[[375,269,4]]_{25}$ \\
$[[21,15,2]]_{49}$ & $[[21,13,3]]_{49}$ & $[[21,11,4]]_{49}$ \\
$[[21,7,5]]_{49}$ & $[[22,8,5]]_{49}$ &$ [[21,5,6]]_{49}$ \\
$[[23,1,7]]_{49}$ & & \\
$[[147,141,2]]_{49}$ &$[[147,127,3]]_{49}$ &$[[147,113,4]]_{49}$ \\
$[[147,85,5]]_{49}$ &
$[[147,71,6]]_{49}$ & \\
\hline

\hline
\caption{Quantum codes obtained from repeated root cyclic codes using Theorem 8} 
\label{tab:ConstAgainstBinaryQECC}
\end{longtable}

\end{document}